# War of 2050: a Battle for Information, Communications, and Computer Security

Alexander Kott, US Army Research Laboratory

David S. Alberts, Institute for Defense Analysis

Cliff Wang, US Army Research Office

**As envisioned in a recent future-casting workshop, warfare will continue to be transformed by advances in information technologies. In fact, information itself will become the decisive domain of warfare.**

-------------------------------------------------

Hopefully there will be no war in 2050 or at any time in the future. But if the past is prologue, humankind's warlike history compels a nation to think about and prepare for the challenges posed by a future conflict.  To the extent we can predict, the bets are that the remarkable surge of information technologies of the last few decades will continue, even accelerate.   Included in this set of technologies is the wide range of information-related and -enabled capabilities that are involved in obtaining, collecting, organizing, fusing, storing, and distributing relevant information as well as the capabilities associated with command and control (C2) functions and processes including reasoning, inference, planning, decision making, and collaborating (between humans, and between humans and machines). Finally, this term includes the capabilities that could be used to deny, deceive, disrupt, degrade, and compromise adversary information and information-related processes (e.g., cyber and electronic warfare).  By the year 2050 these technologies, and the capabilities they offer, will transform the battlefield as we know it.

This motivated the workshop organized by the University of Maryland and sponsored by the US Army Research Office in March 2015.  Over 30 technologists, scientists, military professionals and futurists attempted to visualize the novel features of a hypothetical war of 2050 – fought by hypothetical combatants who do not necessarily include USA – based upon their understandings of technology trends and the application of technology to warfighting. Four defining aspects of the battlefield of 2050 emerged from workshop discussions.  These were: the prevalence of intelligent systems; a force that features enhanced humans; and the decisive battle in the information domain or cyberspace.  Yet another critical challenge for the combatants is to effectively command and control the collection of humans and intelligent systems that will populate the battlefield of the future, in the face of an uncertain and potentially degraded information environment.

## Ubiquitous Intelligent Systems

Whether called "robots" or something else, a variety of intelligent systems, operating with varying and controlled degrees of autonomy, will continue to proliferate. Fueled by steady advances in machine perception and reasoning, intelligent systems, in the view of the workshop's participants, will be a ubiquitous presence on the battlefield of 2050. [1] Sensing, communicating, collaborating intelligent entities will densely populate the battlespace exhibiting a range of sophisticated capabilities. These capabilities will include selectively collecting and processing information, acting as agents to support sensemaking, undertaking coordinated defensive actions, and unleashing a variety of effects on the adversary.

Many of the robots will resemble improved versions of current systems: unattended ground sensors; UAVs; and, fire-and-forget missiles. Some would be robotic physical entities ranging from very small (insect-sized), with bio-inspired locomotion to mid-size (e.g., troop transports) moving over the ground and in the air. Many of them will possess organic sensors and collectively serve as 'sensing fields' providing persistent and complete coverage of the battlefield. Other robots will act as defensive shields, or as intelligent munitions operating alone or in 'wolf packs'.

The latter ones, as well as all systems capable of applying force, will not be autonomous but rather remain under human control. This assumes the hypothetical combatants of 2050 comply with a ban on "offensive autonomous weapons beyond meaningful human control." Such a ban has been called for in a recent open letter by numerous scientists [2], whose concerns about a potential AI arms race the authors of this column fully share. Although the US Department of Defense already imposes strong restrictions on autonomous and semi-autonomous weapon systems [3], positions of other technologically-capable (in 2050) countries is unknowable.

Others robots will be virtual, that is they will be cyberbots that reside within numerous computing systems of the battle, and maneuver and operate in its cyberspace. They will: protect communications and information; defend electronic devices of the humans; prevent or warn about incoming threats; and, advise decision makers. In addition, these cyberbots could potentially carry out proactive actions such as to disguise friendly forces' presence in both physical and cyber space, penetrate adversary systems and deceive adversaries into false observations and wrong decisions.

To perform these roles, battlefield robots will be robustly networked, dynamically interacting with each other and their human controllers, clients, and collaborators. They will, when required, self-organize. They will employ a variety of control modes from total autonomy to being actively managed by humans under dynamically established rules of engagement and priorities. Decision agents would be integral to all of the processes of commanding the forces and gathering battlefield intelligence. These cyberspace robots will fact-check, filter and fuse information [4]. They will determine who has access to what information and will disseminate information adaptively. In addition, they will route communications, assign tasks to sensors, and perform coordinated actions with physical bots.

Executing so many critical functions, robots will be valuable targets—and will present a large surface—for the adversary's attacks. Robots will be particularly vulnerable to various types of attacks targeting their information, information processing and communications, in addition to being subjected to physical attack or capture. Therefore, computer and network security will be a paramount consideration in the design and operation of robots, both physical and cyber, and the networks that support the human-robot teams populating the battlefield of 2050.

## Human Warriors with Superhuman Abilities

The battlefield of the future will be populated by fewer humans. But these warriors will be physically and cognitively augmented. [5] This will improve their ability to: sense; make sense; and interact with each other humans, and with automated processes.  As a result, they would not only do things differently but do different things.  Human-robot teams will be the principal units of military forces operating in 2050. For this reason, augmentation of human abilities will be important in order to partner effectively with robots and to function effectively in an information rich environment.  Augmented humans will be enabled by seamless access to sensing and computing power extending the trend toward more natural man-machine integration. Although seemingly unlikely, one round of a remarkable human augmentation has already occurred: the ubiquitous smart phone, nearly inseparable from our persona, has extended humans' abilities to obtain, process and communicate information to a dramatic degree, unimaginable merely 30 years ago.

While workshop participants did not expect that in 2050, the information 'wheat' could be readily extracted from the chaff, they felt that individuals' (and organizations') would have much improved abilities to cope with imperfect information and  more accurately understand the limitations and risks associated with the available information. This result was expected for four reasons.  First, 2050 warriors will be digital natives and, of necessity, will have developed information survival skills; second, their cognitive capabilities will have been enhanced; third, they will be assisted by cyberbots to help in verify sources; and, fourth, have improved visualization and other human-machine interfaces. In addition to these super-human cognitive capabilities, 2050-era warriors may feature exoskeletons that augment their physical powers.

Workshop participants noted that with so many human enhancements dependent on computing devices, the enhanced human will become subject—like robots and cyberbots—to a variety of information attacks. These will include denial of service attacks, hacking, spoofing and electronic warfare in efforts to compromise embedded computers and the network, to prevent access to processing power, information sources, and collaborations. It is critical to create attack resilient, mission sustaining cyber systems that can support augmented warriors to survive extremely tough battle field environments that consist of both kinetic force damages as well as intelligent cyber attacks.

## Decision Battle for the Information Domain

With so many smart, sharp-eyed things blanketing the battlefield and reporting their detailed observations, it will become increasingly difficult to hide one's forces from the adversary. Efforts to cloak/uncloak assets will be intense.  Disinformation and deception will be essential to survive and operate on the battlefield of 2050.   Workshop participants concluded that the ability to extract value from information while preventing adversaries from doing the same will become the decisive factor in the War of 2050.

This development is a direct result of the transition from Industrial Age to Information Age Warfare. [6] [7]  Until quite recently, the only information a soldier received was from a few authoritative and trusted sources. As information became separated from the chain of command, soldiers began to have access to more information sources but a new problem has arisen—assessing the quality and trustworthiness of information.

By 2050, the highly developed science and technology of synthesizing believable misinformation—and delivering it to the adversary through a variety of network and malware channels—will make it difficult to assess the quality, correctness, authenticity, and security of information.   Misinformation attacks will be hard to detect and when undetected will sow mis-trust and confusion, and delay and undermine decision making.[10] The battle for information, and against the mis-information, will become qualitatively more acute and critical than ever before in the history of warfare. [8]

# Command and Control: A Critical Challenge

The war of 2050 will feature a crowded battlefield where physical entities have fewer places to hide and where cyber entities will be difficult to detect and track. The level of trust in information and the availability of communications will be severely tested by the intense battle for the information domain. While operating on a crowded battlefield is certainly complicated, the increased transparency of physical entities, the potential of adversaries to wreak havoc in the information domain, and the increased degree of freedom granted to some of intelligent entities will transform the command and control problem from a complicated one to a complex one. This has profound implications for approaches to command and control, and for the related systems.

Traditional approaches to command and control have evolved a set of variants of a hierarchical approach that have proven to work well for complicated situations. In fact, militaries are widely respected for their ability to manage these situations. Recent experience and research show that these traditional approaches to command and control are less well-adapted for dealing with the complexity that will be a feature of the battlefield of 2050. Thus, the combatants of 2050 will have two meet two critical command and control-related challenges: 1) successfully manage the teams of forces that would act, as the situation and policy requires, independently or collectively and 2) manage and protect the communications and information networks that enable effective management.

Workshops participants envisioned a battlefield where forces are required to self-organize on a large scale and where collective decision-making is the norm. This more collaborative, networked approach to command and control was seen as critical to enable a heterogeneous collection of human-machine teams to cope in the highly contested, imperfect information environment that will characterize the battlefield of 2050.

# Conclusion

While the War of 2050 will still be a human dominated affair, four developments will significantly change the nature of the battle. [9] The first of these will be a proliferation of intelligent systems; the second, augmented humans; the third, the decisive battle for the information domain; and the fourth, the introduction of new, networked approaches to command and control. Each of these new capabilities possesses the same critical vulnerability – attacks on the information, communications and computers that will enable human-robot teams to make sense of the battlefield and act decisively. Hence, the largely unseen battle for information, communications and computer security will determine the extent to which adversaries will be able to function and succeed on the battlefield of 2050.

# Disclaimer

All views and opinions of this article belong to the authors and do not represent those of their employers.